\documentclass[prd,tightenlines]{revtex4}
\usepackage{amssymb}

\usepackage{amsmath}
\usepackage[dvips,final]{graphicx}
\usepackage{epsfig}
\usepackage{bm}

\begin{document}

\date{February 8, 2010}
\title{Comments on "Single meson production in photon-photon collisions and
infrared renormalons"[arXiv:0911.5226v1][hep-ph] \\
}
\author{\textbf{S.~S.~Agaev}}
\email{agaev_shahin@yahoo.com}
\affiliation{Institute for Physical Problems, Baku State University\\
Z. Khalilov St.\ 23, Az-1148 Baku, Azerbaijan}

\begin{abstract}
This paper contains our comments on the work of the authors
A.~I.~Ahmadov \textit{et al.} "Single meson production in
photon-photon collisions and infrared renormalons"
[arXiv:0911.5226v1 [hep-ph] 27 Nov 2009]. We draw attention to
errors made in this work in calculation of the subprocess cross
sections within the frozen and running coupling approaches, in
deriving the Borel transforms and Borel resummed expressions for
the cross sections in the running coupling approach. We show that
the same fatal errors were made by these authors also in their
work published in Phys. Rev. D80, 016003 (2009).
\end{abstract}

\maketitle

In the work ''Single meson production in photon-photon collisions and
infrared renormalons'' [arXiv:0911.5226v1 [hep-ph] 27 Nov 2009] \cite{Az1}
authors A.~I. Ahmadov, Coskun Aydin, E. A. Dadashov and Sh. M. Nagiyev in
accordance with the title of the paper intended to compute the cross section
of the single meson $M$ production in the photon-photon collision $\gamma
+\gamma \rightarrow M+X$.

In the framework of the perturbative QCD (pQCD) this process proceeds
through leading and higher twist mechanisms. At the leading twist (LT) order
the meson $M$ is produced due to fragmentation of the final quark or gluon
(appearing in the hard-scattering subprocess) into the meson $M$. The
relevant LT cross section can be computed using the standard pQCD methods,
applicable to inclusive hadronic processes. The higher twist (HT) mechanism
implies production of the meson $M$ directly at the hard-scattering
subprocess \cite{Br01}. In the HT subprocess emergence of the meson in the
final state becomes possible in the result of the hard gluon exchange
between meson constituents. The cross section of the meson production via HT
mechanism can be found employing the methods of the pQCD elaborated to
compute hard exclusive processes and relevant factorization theorems \cite%
{Br02,Rad80,Muel80}. By this way HT corrections to some inclusive processes
were obtained \cite{Bar80,Bag,Has,Ag01}.

In order to find the amplitude \ of the HT hard-scattering subprocess one
should use the distribution amplitude of the meson and perform integrations
over the longitudinal momentum fractions $x_{1}$ and $x_{2}$ carried by
quark and antiquark of the meson. In the case under discussion it takes the
following form%
\begin{equation}
M=\int_{0}^{1}\int_{0}^{1}dx_{1}dx_{2}\delta (1-x_{1}-x_{2})\alpha _{s}(\mu
_{R}^{2})T_{H}(x_{1},x_{2},\mu _{F}^{2})\Phi _{M}(x_{1},x_{2},\mu _{F}^{2}),
\label{eq:1}
\end{equation}%
where $\mu _{R}^{2}$ and $\mu _{F}^{2}$ are the renormalization
and factorization scales, $T_{H}(x_{1},x_{2},\mu _{F}^{2})$ -
hard-scattering function, $\Phi _{M}(x_{1},x_{2},\mu _{F}^{2})$ is
the distribution amplitude of the meson. One of the important
questions to be solved in order to calculate Eq.(\ref{eq:1}) is a
proper choice for the scales $\mu _{R}^{2}$ and $\mu _{F}^{2}$. It
has been advocated \cite{Bag, Br03} that to reduce the higher
order corrections to a physical quantity and improve the
convergence of the corresponding perturbation series, the
renormalization scale, i.e., the argument of the QCD coupling in a
Feynman diagram should be set equal to the hard gluon's squared
four-momentum. This idea was proposed more than 25 years ago, and
used in numerous works. Nevertheless, A.I. Ahmadov \textit{et al.}
write: ''According to Ref.[26] [A.I.Ahmadov \textit{et al.}, Int.
J. Mod. Phys. E15, 1209 (2006)] should be noted that in PQCD
calculations the argument of the running coupling constant in
both, the renormalization and factorization scale $\hat{Q}^{2}$
should be taken equal to the square of the momentum transfer of a
hard gluon in a corresponding Feynman diagram''!

In exclusive processes and in HT Feynman diagrams describing corrections to
inclusive ones, the renormalization scale chosen this way inevitably depends
on the longitudinal momentum fractions $x_{1}$ and $x_{2}$ carried by the
meson's quark and/or antiquark $\mu _{R}^{2}\thicksim x_{1}(x_{2})$. Then
the pQCD factorization formula (in our case Eq.(\ref{eq:1})) diverges, since
$\alpha _{s}(\mu _{R}^{2})$ suffers from an end-point $x_{1}(x_{2})%
\rightarrow 0,1$ singularities. Therefore, the HT amplitude may be
computed using two options: one of them is the standard ''frozen''
coupling approximation, when one fixes $x_{1}$ and $x_{2}$
equating them to their mean values $x_{1}=x_{2}=1/2$, and remove
$\alpha _{s}(\mu _{R}^{2})$ as \ a constant from the integral
Eq.(\ref{eq:1}). In the second approach one allows $x_{1}$ and
$x_{2}$ to run in the argument of the $\alpha _{s}(\mu _{R}^{2})$
and calculate the HT amplitude using the running coupling (RC)
method, and removes divergences appearing in the perturbative
expression with the help of the infrared (IR) renormalon calculus
\cite{Beneke} (the Borel transformation, resummed expression for a
physical quantity) and the principal value prescription
\cite{Sterman}. It turns out that this method allows one to
estimate power corrections arising from the end-point integration
regions.

The RC method was suggested in our works \cite{Ag02,Ag03} (see also, \cite%
{stef}) in order to compute effects of the infrared renormalons on
the pion and kaon electromagnetic form factors (FFs). Later the RC
method was applied for calculation of power suppressed corrections
to some exclusive processes, vertex function and HT corrections to
semi-inclusive processes.
Namely, it was used for computations of the pion, kaon electromagnetic FFs %
\cite{Ag02,Ag03,Agaev1}, $\pi ^{0}\gamma $ \cite{Ag04} and $\eta \gamma
,\,\eta \prime \gamma $ electromagnetic transition FFs \cite{Ag05}, for
evaluation of the power corrections to the gluon-gluon-$\eta ^{\prime }$
vertex function \cite{Ag06} and to the two pion production in the process $%
\gamma \gamma ^{\ast }\rightarrow \pi \pi $ \cite{Ag07}. It was also
employed to estimate HT correction to the semi-inclusive meson production $%
\gamma +h\rightarrow M+X$ \cite{Ag08}. It is worth noting that the RC method
deals with power-suppressed corrections to exclusive processes coming from
the end-point integration regions. There are another source of the infrared
renormalon effects in inclusive and exclusive processes, namely one
appearing due to resummation of diagrams with quark vacuum insertions
(''bubble chains'') into a gluon line. It is known that resummation of
infinite number of such diagrams corresponds to computation of one-loop
diagrams with the runnning coupling $\alpha _{s}(-k^{2})$ \ at vertices. The
HT ambiguities produced during these calculations were employed to model HT
corrections to numerous inclusive and exclusive processes (for review, see %
\cite{Beneke}). In the exclusive processes the coupling constant $\alpha
_{s} $ runs not only due to loop integration, but also because of the
integration in the process amplitude over the longitudinal momentum
fractions of hadron constituents. Thus the exclusive processes have two
independent sources of the infrared renormalon effects.

In Ref. \cite{Az1} both the frozen and running coupling approaches were
used. Employing the RC method elaborated in Refs. \cite%
{Ag02,Ag03,Ag04,Ag05,Ag06,Ag07,Ag08}, authors of \cite{Az1} nevertheless
''forgot'' to cite them - the real sources used in their calculations and in
preparation of the text of the paper. Now we want to concentrate on
''calculations'' performed in Ref. \cite{Az1}. The higher twist subprocesses
considered in Ref. \cite{Az1} are%
\begin{equation}
\gamma q_{1}\rightarrow Mq_{2},\;\ \gamma \overline{q}_{2}\rightarrow M%
\overline{q}_{1},\;\;\ \ M=(q_{1}\overline{q}_{2}).  \label{eq:1a}
\end{equation}%
In the ''frozen'' coupling approximation they were calculated by Bagger and
Gunion \cite{Bag}. The authors presented their results for the $\gamma
q_{1}\rightarrow Mq_{2}$ subprocess cross section in Eqs. (16) and (17) of
Ref. \cite{Bag}:%
\begin{equation*}
\frac{d\sigma }{d\hat{t}}\mid _{\gamma q\rightarrow Mq}(\hat{s},\hat{t},%
\hat{u})=\frac{8\pi ^{2}C_{F}\alpha _{E}}{9}\left[ \Delta (\hat{s},\hat{u})%
\right] ^{2}\frac{1}{\hat{s}^{2}(-\hat{t})}\left[ \frac{1}{\hat{s}^{2}}+%
\frac{1}{\hat{u}^{2}}\right] ,\;\;\;\;\;\;\pi ,\,\rho _{L},
\end{equation*}%
\begin{equation}
\frac{d\sigma }{d\hat{t}}\mid _{\gamma q\rightarrow Mq}(\hat{s},\hat{t},%
\hat{u})=\frac{8\pi ^{2}C_{F}\alpha _{E}}{9}\left[ \Delta (\hat{s},\hat{u})%
\right] ^{2}\frac{8(-\hat{t})}{\hat{s}^{4}\hat{u}^{2}},\;\;\;\rho _{T},
\label{eq:1b}
\end{equation}%
where%
\begin{equation}
\Delta (\hat{s},\hat{u})=\left[ \hat{u}e_{1}\alpha _{s}\left[ \frac{\hat{s}}{%
2}\right] I_{M}\left[ \frac{\hat{s}}{2}\right] +\hat{s}e_{2}\alpha _{s}\left[
-\frac{\hat{u}}{2}\right] I_{M}\left[ -\frac{\hat{u}}{2}\right] \right] .
\label{eq:1c}
\end{equation}%
In Eqs.(\ref{eq:1b}) and (\ref{eq:1c}) $\hat{s},\;\hat{u},\;\hat{t}$ are the
Mandelstam invariants of the subprocess, $e_{1}$ and $e_{2}$ are the charges
of the quark $q_{1}$ and antiquark $\overline{q}_{2}.$ The function $I_{M}$
in Ref. \cite{Bag} is defined as%
\begin{equation}
I_{M}("Q^{2}")=\int_{0}^{1}\frac{dx\Phi _{M}(x,"Q^{2}")}{x(1-x)},
\label{eq:1d}
\end{equation}%
where $"Q^{2}"=\shortmid t_{g}\shortmid $ and ''$t_{g}$ is the
average squared momentum transfer carried by the hard gluon in a
given subprocess'' equal to $\hat{s}/2$ and $-\hat{u}/2$ for two
type of relevant Feynman diagrams.

In our work ''Single meson photoproduction and IR renormalons'' \cite{Ag08}
it was demonstrated that namely the HT subprocesses (\ref{eq:1a}) contribute
to the photoproduction of the single meson $\gamma +h\rightarrow M+X$ via
the HT mechanism. Within the RC method the differential cross sections of
these HT subprocesses were found in our work \cite{Ag08}. The differential
cross section of the subprocess $\gamma q_{1}\rightarrow Mq_{2}$, when $M$
is a pseudoscalar or longitudinally polarized vector meson, is given by the
following expression [Eq.(9) in Ref. \cite{Ag08}]:

\begin{equation*}
\frac{d\hat{\sigma}^{HT}(e_{1},e_{2})}{d\hat{t}}=\frac{32\pi ^{2}C_{F}\alpha
_{E}}{9\hat{s}^{2}}\left\{ -\frac{e_{1}^{2}}{\hat{s}^{2}}\left[ I_{1}^{2}%
\hat{t}-2I_{1}\left( I_{1}\hat{s}+I_{2}\hat{u}\right) \frac{\hat{u}}{\hat{t}}%
+I_{2}^{2}\frac{\hat{u}^{2}}{\hat{t}}\right] -\frac{e_{2}^{2}}{\hat{u}^{2}}%
\left[ K_{1}^{2}\hat{t}-2K_{1}\left( K_{1}\hat{u}+K_{2}\hat{s}\right) \frac{%
\hat{s}}{\hat{t}}+K_{2}^{2}\frac{\hat{s}^{2}}{\hat{t}}\right] \right.
\end{equation*}%
\begin{equation}
\left. -\frac{2e_{1}e_{2}}{\hat{s}\hat{u}\hat{t}}\left[ I_{1}K_{1}\hat{t}%
^{2}-I_{1}\left( K_{2}\hat{s}+K_{1}\hat{u}\right) \hat{s}-K_{1}\left( I_{1}%
\hat{s}+I_{2}\hat{u}\right) \hat{u}\right] \right\} .  \label{eq:2}
\end{equation}%
For the transversely polarized vector meson we get [Eq.(10) in
Ref. \cite{Ag08}]:

\begin{equation}
\frac{d\hat{\sigma}^{HT}(e_{1},e_{2})}{d\hat{t}}=\frac{64\pi ^{2}C_{F}\alpha
_{E}}{9\hat{s}^{4}}\frac{-\hat{t}}{\hat{u}^{2}}\left[ e_{1}\hat{u}I_{2}-e_{2}%
\hat{s}K_{2}\right] ^{2}.  \label{eq:3}
\end{equation}%
In Eqs.(\ref{eq:2}),(\ref{eq:3}) the functions $I_{1(2)}$ and
$K_{1(2)}$ encode an information on the meson distribution
amplitude and have the forms:

\begin{equation}
I_{1}=\int_{0}^{1}\int_{0}^{1}\frac{dx_{1}dx_{2}\delta (1-x_{1}-x_{2})\alpha
_{s}(\mu _{R1}^{2})\Phi _{M}(x_{1},x_{2}; \mu _{F1}^{2})}{x_{2}}%
,\;I_{2}=\int_{0}^{1}\int_{0}^{1}\frac{dx_{1}dx_{2}\delta
(1-x_{1}-x_{2})\alpha _{s}(\mu _{R1}^{2})\Phi _{M}(x_{1},x_{2}; \mu _{F1}^{2})%
}{x_{1}x_{2}},  \label{eq:4}
\end{equation}

\begin{equation}
K_{1}=\int_{0}^{1}\int_{0}^{1}\frac{dx_{1}dx_{2}\delta (1-x_{1}-x_{2})\alpha
_{s}(\mu _{R2}^{2})\Phi _{M}(x_{1},x_{2}; \mu _{F2}^{2})}{x_{1}}%
,\;K_{2}=\int_{0}^{1}\int_{0}^{1}\frac{dx_{1}dx_{2}\delta
(1-x_{1}-x_{2})\alpha _{s}(\mu _{R2}^{2})\Phi _{M}(x_{1},x_{2}; \mu _{F2}^{2})%
}{x_{1}x_{2}}.  \label{eq:5}
\end{equation}%
Here
\begin{equation*}
\mu _{R1}^{2}=x_{2}\hat{s},\;\;\mu _{F1}^{2}=\hat{s}/2;\;\;\mu
_{R2}^{2}=-x_{1}\hat{u},\;\mu _{F2}^{2}=-\hat{u}/2,
\end{equation*}
$\hat{s},\;\hat{u},\;\hat{t}$ are the Mandelstam invariants of the
subprocess, $e_{1}$ and $e_{2}$ are the charges of the quarks $q_{1}$ and $%
q_{2}$, respectively.

Equations (\ref{eq:2}), (\ref{eq:3}) are general expressions valid for
mesons with both the symmetric (under exchange $x_{1}\leftrightarrow x_{2}$)
and non-symmetric distribution amplitudes. In the frozen coupling
approximation, i.e., when
\begin{equation*}
\mu _{R1}^{2}=\hat{s}/2;\;\;\mu _{R2}^{2}=-\hat{u}/2,
\end{equation*}%
from Eqs. (\ref{eq:2}), (\ref{eq:3}) for mesons with symmetric distributions
the Bagger-Gunion results can be obtained. Indeed, in this approximation for
the functions $I_{1(2)}$ and $K_{1(2)}$ we find

\begin{equation*}
I_{2}^{0}\left( \frac{\hat{s}}{2}\right) =2I_{1}^{0}\left( \frac{\hat{s}}{2}%
\right) =\alpha _{s}\left( \frac{\hat{s}}{2}\right) \int_{0}^{1}\frac{dx\Phi
_{M}(x,\hat{s}/2)}{x(1-x)},
\end{equation*}%
\begin{equation}
K_{2}^{0}\left( -\frac{\hat{u}}{2}\right) =2K_{1}^{0}\left( -\frac{\hat{u}}{2%
}\right) =\alpha _{s}\left( -\frac{\hat{u}}{2}\right) \int_{0}^{1}\frac{%
dx\Phi _{M}(x,-\hat{u}/2)}{x(1-x)}.  \label{eq:5a}
\end{equation}%
Here the superscript ''$0$'' indicates that the functions are found in the
frozen coupling approximation. Having performed some analytical calculations
we get%
\begin{equation}
\frac{d\hat{\sigma}^{HT}(e_{1},e_{2})}{d\hat{t}}=\frac{8\pi ^{2}C_{F}\alpha
_{E}}{9\hat{s}^{2}}\frac{1}{(-\hat{t})}\left[ e_{1}\hat{u}I_{2}^{0}-e_{2}%
\hat{s}K_{2}^{0}\right] ^{2}\left[ \frac{1}{\hat{s}^{2}}+\frac{1}{\hat{u}^{2}%
}\right] ,  \label{eq:6}
\end{equation}

\begin{equation}
\frac{d\hat{\sigma}^{HT}(e_{1},e_{2})}{d\hat{t}}=\frac{64\pi ^{2}C_{F}\alpha
_{E}}{9\hat{s}^{4}}\frac{-\hat{t}}{\hat{u}^{2}}\left[ e_{1}\hat{u}%
I_{2}^{0}-e_{2}\hat{s}K_{2}^{0}\right] ^{2}.  \label{eq:7}
\end{equation}%
These expressions coincide with the Bagger-Gunion results, if to take into
account that in Ref. \cite{Bag} $e_{1}$ and $e_{2}$ are the charges of the
quark $q_{1}$ and antiquark $\overline{q}_{2}$, whereas in our work they
denote the charges of the quarks $q_{1}$ and $q_{2}$.

In Ref. \cite{Az1} authors presented results of their calculations
of the subprocess $\gamma q_{1}\rightarrow Mq_{2}$ cross section.
In accordance with Eqs.(2.10) and (2.11) of Ref. \cite{Az1} these
cross sections are given
by the following expressions:%
\begin{equation*}
\frac{d\sigma }{d\hat{t}}(\gamma q\rightarrow Mq)=\frac{8\pi C_{F}\alpha _{E}%
}{9}\left[ D(\hat{s},\hat{u})\right] ^{2}\frac{1}{\hat{s}^{2}(-\hat{t})}%
\left[ \frac{1}{\hat{s}^{2}}+\frac{1}{\hat{u}^{2}}\right] ,\;\;\;\;\;M=\;\pi
,\,\rho _{L},
\end{equation*}%
\begin{equation}
\frac{d\sigma }{d\hat{t}}(\gamma q\rightarrow Mq)=\frac{8\pi C_{F}\alpha _{E}%
}{9}\left[ D(\hat{s},\hat{u})\right] ^{2}\frac{8(-\hat{t})}{\hat{s}^{4}%
\hat{u}^{2}}\left[ \frac{1}{\hat{s}^{2}}+\frac{1}{\hat{u}^{2}}\right]
,\;\;\;M=\rho _{T},  \label{eq:8}
\end{equation}%
where%
\begin{equation}
D(\hat{t},\hat{u})=e_{1}\hat{t}\int_{0}^{1}dx_{1}\left[ \frac{\alpha
_{s}(Q_{1}^{2})\Phi _{M}(x,Q_{1}^{2})}{1-x_{1}}\right] +e_{2}\hat{u}%
\int_{0}^{1}dx_{1}\left[ \frac{\alpha _{s}(Q_{2}^{2})\Phi _{M}(x,Q_{2}^{2})}{%
1-x_{1}}\right]  \label{eq:9}
\end{equation}%
and $Q_{1}^{2}=\hat{s}/2$, $Q_{2}^{2}=-\hat{u}/2.$

Now using Eq.(\ref{eq:9}) [Eq.(2.11) in Ref.\cite{Az1}] for the function $D(%
\hat{s},\hat{u})$ we find
\begin{equation}
D(\hat{s},\hat{u})=e_{1}\hat{s}\int_{0}^{1}dx_{1}\left[ \frac{\alpha _{s}(%
\hat{s}/2)\Phi _{M}(x,\hat{s}/2)}{1-x_{1}}\right] +e_{2}\hat{u}%
\int_{0}^{1}dx_{1}\left[ \frac{\alpha _{s}(-\hat{u}/2)\Phi _{M}(x,-\hat{u}/2)%
}{1-x_{1}}\right]  \label{eq:10}
\end{equation}%
It is evident that Eqs.(\ref{eq:8}) and (\ref{eq:10}) are the subprocess
cross sections computed in the framework of the frozen coupling
approximation. Having compared them with Bagger-Gunion expressions (\ref%
{eq:1b}) and (\ref{eq:1c}) one can fix errors made in
Ref.\cite{Az1}. Namely, ignoring error/misprint (instead of $\pi$
should be $\pi ^{2}$) we
see that the function $D(\hat{s},\hat{u})$ does not coincide with $\Delta (%
\hat{s},\hat{u})$ from Eq.(\ref{eq:1c}): dependence on the subprocess
invariants $\hat{s}$ and $\hat{u}$ are wrong, and normalization is also
incorrect, because for the symmetric distribution amplitudes
\begin{equation*}
I_{M}(\hat{s}/2)=\int_{0}^{1}\frac{dx\Phi _{M}(x,\hat{s}/2)}{x(1-x)}%
=\int_{0}^{1}\frac{dx\Phi _{M}(x,\hat{s}/2)}{(1-x)}+\int_{0}^{1}\frac{dx\Phi
_{M}(x,\hat{s}/2)}{x}=2\int_{0}^{1}dx_{1}\left[ \frac{\Phi _{M}(x,\hat{s}/2)%
}{1-x_{1}}\right] .
\end{equation*}%
In other words, the results of Ref. \cite{Az1} contain additional numerical
factor $1/4$.

In Sec. III authors tried to compute the subprocess cross section in the RC
method. Explaining the choice of the renormalization and factorization
scales made in their work, they write: ''As is seen from (2.11) [Eq.( \ref%
{eq:9})], in general, one has to take into account not only dependence of $%
\alpha (\hat{Q}_{1,2}^{2})$ on the scale $\hat{Q}_{1,2}^{2}$, but also an
evolution of $\Phi (x,\hat{Q}_{1,2}^{2})$ with $\hat{Q}_{1,2}^{2}$. The
meson wave function evolves in accordance with a Bete-Salpeter-type
equation. Therefore, it is worth noting that, the renormalization scale
(argument of $\alpha _{s}$) should be equal to $Q_{1}^{2}=-x\hat{u}%
,\;Q_{2}^{2}=\hat{s}x$, whereas the factorization scale [$Q^{2}$
in $\Phi (x,Q^{2}$]$\;$is taken independent from $x$, we take
$Q^{2}=p_{T}^{2}$''. In the frozen coupling approximation, as is
evident also from the Bagger-Gunion
expression (\ref{eq:1c}), the choice for the factorization scales is $\hat{s}%
/2$ and $-\hat{u}/2$, whereas in the RC method without any reasons authors
put $\mu _{F1}^{2}=\mu _{F2}^{2}=p_{T}^{2}$ , $p_{T}^{2}$ being the meson $M$
transverse momentum square. Of course, such choice is not legitimate in the
context of RC method. Actually authors did not compute the subprocess cross
section using the RC method, but ''generalized'' their wrong expression
obtained in the frozen coupling approximation by this way [the first line in
Eq.(3.5) of Ref.\cite{Az1}]:%
\begin{equation}
D(Q^{2})=e_{1}\hat{t}\int_{0}^{1}dx\frac{\alpha _{s}(\lambda Q^{2})\Phi
_{M}(x,Q^{2})}{1-x}+e_{2}\hat{u}\int_{0}^{1}dx\frac{\alpha _{s}(\lambda
Q^{2})\Phi _{M}(x,Q^{2})}{1-x},  \label{eq:11}
\end{equation}%
where $\lambda =1-x$ (as it follows later from Eq. (3.7)).

In our work \cite{Ag08} it was proved that within the RC method
the cross section of the subprocess $\gamma q_{1}\rightarrow
Mq_{2}$ is given by Eqs.( \ref{eq:2}), (\ref{eq:3}). Simple
generalization of the Bagger-Gunion formula (\ref{eq:1b}) for the
pseudoscalar and longitudinally polarized vector meson in order to
use it in the RC method leads to wrong results, because in the RC
method even for mesons with symmetric distribution amplitudes
relations (\ref{eq:5a}) do not hold. The latter conclusion was
made in Ref. \cite{Ag08} and proved by explicit calculations,
results of which were presented in Eqs. (30) and (32) of this
paper.

As is seen from Eq.(\ref{eq:11}), in Ref. \cite{Az1} even for the
renormalization scales $\mu _{R1}^{2},\;\mu _{R2}^{2}$ the choice
$\mu _{R1}^{2}=\mu _{R2}^{2}=(1-x)p_{T}^{2}$ was made, which
contradicts to principles of the RC method and declarations of the
authors. Therefore, Eqs.(3.9)-(3.13) are wrong themselves and do
not describe situation with the two scales. The Borel transforms
presented in Eqs.(3.14)-(3.17) and resummed expressions shown in
Eqs.(3.18)-(3.21) are also wrong in the case of two scales. We
think that we can stop at this point our analysis of the
''investigation'' \ performed in Ref.\cite{Az1} .

Unfortunately, exactly the same errors were made by these authors
in their paper published in Phys. Rev. D80, 016003 (2009) and
entitled ''Infrared
renormalons and single meson production in proton-proton collisions'' \cite%
{Az2}. Despite declarations of the authors the process that they tried to
considere is not ''single meson production''%
\begin{equation*}
p+p\rightarrow M+X,
\end{equation*}%
but ''prompt photon and meson production''%
\begin{equation*}
p+p\rightarrow \gamma +M+X.
\end{equation*}

In Eq.(2.10) of Ref. \cite{Az2} authors present the diffential cross section
of the HT subprocess $q_{1}\overline{q}_{2}\rightarrow \pi ^{+}(\pi
^{-})+\gamma $%
\begin{equation}
\frac{d\sigma }{d\hat{t}}(\hat{s},\hat{t},\hat{u})=\frac{8\pi
^{2}C_{F}\alpha _{E}}{27}\left[ D(\hat{t},\hat{u})\right] ^{2}\frac{1}{%
\hat{s}^{3}}\left[ \frac{1}{\hat{t}^{2}}+\frac{1}{\hat{u}^{2}}\right] ,
\label{eq:12}
\end{equation}%
where%
\begin{equation}
D(\hat{t},\hat{u})=e_{1}\hat{t}\int_{0}^{1}dx_{1}\left[ \frac{\alpha
_{s}(Q_{1}^{2})\Phi _{M}(x,Q_{1}^{2})}{1-x_{1}}\right] +e_{2}\hat{u}%
\int_{0}^{1}dx_{1}\left[ \frac{\alpha _{s}(Q_{2}^{2})\Phi _{M}(x,Q_{2}^{2})}{%
1-x_{1}}\right] .  \label{eq:13}
\end{equation}%
Here $Q_{1}^{2}=-(1-x_{1})\hat{u}$ and $Q_{2}^{2}=-x_{1}\hat{t}$.

In Sec.III of Ref. \cite{Az2} authors tried to compute the subprocess $q_{1}%
\overline{q}_{2}\rightarrow \pi ^{+}(\pi ^{-})+\gamma $ cross section in the
context of the RC method. But instead of calculation of the Feynman diagrams
of the subprocess within the RC method they ''generalize'' the expression (%
\ref{eq:12}) obtained in the frozen coupling approach. The correct
expression for the subprocess cross section in the RC approach can be
obtained from our formula (\ref{eq:2}) for the pseudoscalar mesons, after
exchange of the Mandelstam invariants and by taking into account numerical
factors.

In this section the authors write:''As is seen from (2.11) [Eq.(\ref{eq:13}%
)], in general, one has to take into account not only dependence of $\alpha (%
\hat{Q}_{1,2}^{2})$ on the scale $\hat{Q}_{1,2}^{2}$, but also an evolution
of $\Phi (x,\hat{Q}_{1,2}^{2})$ with $\hat{Q}_{1,2}^{2}$. The meson wave
function evolves in accordance with a Bete-Salpeter-type equation.
Therefore, it is worth noting that, the renormalization scale (argument of $%
\alpha _{s}$) should be equal to $Q_{1}^{2}=(x_{1}-1)\hat{u},\;Q_{2}^{2}=-%
\hat{t}x_{1}$, whereas the factorization scale [$Q^{2}$ in $\Phi (x,Q^{2}$]$%
\;$is taken independently from $x$; we take $Q^{2}=p_{T}^{2}$ ''. Such
choice for the factorization scales, i.e. equating them to the pion's
transverse momentum square $p_{T}^{2}$, in the HT subprocess calculations is
wrong. The factorization scales had to be chosen equal to $-\hat{u}/2$ and $-%
\hat{t}/2$ in accordance with the standard prescriptions. As is
seen from Eq.(3.5) of \cite{Az2} and the text above of the
Eq.(3.7), not only the factorizations scales, but also the
renormalization scales were chosen in a way that contradicts
prescriptions of the RC method and statements of the authors.
Indeed, in the first line of Eq.(3.5) having
corrected the misprint (should be $x$ instead of $x_{1}$) we find%
\begin{equation}
D(Q^{2})=e_{1}\hat{t}\int_{0}^{1}dx\frac{\alpha _{s}(\lambda Q^{2})\Phi
_{M}(x,Q^{2})}{1-x}+e_{2}\hat{u}\int_{0}^{1}dx\frac{\alpha _{s}(\lambda
Q^{2})\Phi _{M}(x,Q^{2})}{1-x},  \label{eq:14}
\end{equation}%
where $\lambda =1-x$. It is evident that in this case $\mu _{R1}^{2}=\mu
_{R2}^{2}=(1-x)p_{T}^{2}$, and they are not equal to $Q_{1}^{2}=(x_{1}-1)%
\hat{u},\;Q_{2}^{2}=-\hat{t}x_{1}$ as write authors in the paper.
In other words, in Ref. \cite{Az2} the factorization and
renormalization scales were expressed using the transverse
momentum square of the \textbf{final particle}, but not the
momentum transfer carried by \textbf{hard gluon}. All computations
in Ref.\cite{Az2} were carried out with these incorrect choices
for both the factorization and renormalization scales, which led
to wrong expressions.

Ignoring for the moment that expression for the function $D(Q^{2})$ (\ref%
{eq:14}) in the RC method is wrong and does not describe pion production,
let us nevertheless explane the correct approach to computation of such
integrals. After correct choices for the scales it should have the form:%
\begin{equation}
D(\hat{t},\hat{u})=e_{1}\hat{t}\int_{0}^{1}dx\frac{\alpha _{s}[-(1-x)\hat{u}%
]\Phi _{M}(x,-\hat{u}/2)}{1-x}+e_{2}\hat{u}\int_{0}^{1}dx\frac{\alpha _{s}[-x%
\hat{t}]\Phi _{M}(x,-\hat{t}/2)}{1-x}.  \label{eq:15}
\end{equation}%
Having expressed the running couplings $\alpha _{s}[-(1-x)\hat{u}]$ and $%
\alpha _{s}[-x\hat{t}]$ in terms of $\alpha _{s}(-\hat{u})$, $\alpha _{s}(-%
\hat{t})$ with the aid of the renormalization group equation we find
\begin{equation*}
\alpha (\lambda Q^{2})=\frac{\alpha (Q^{2})}{1+\ln \lambda /t}
\end{equation*}%
where $\alpha =\alpha _{s}/\pi $ and $t=4\pi /\beta _{0}\alpha _{s}(Q^{2})$
(there is an error/misprint in corresponding Eq.(3.4) of Ref.\cite{Az2}).
Having used this expression the integrals in Eq. (\ref{eq:15}) can be
recasted into the following forms:%
\begin{equation}
D(\hat{t},\hat{u})=\frac{4\sqrt{3}\pi f_{\pi }e_{1}\hat{t}}{\beta _{0}}%
\int_{0}^{\infty }due^{-t_{1}u}A[u,\alpha _{s}(-\hat{u}/2)]+\frac{4\sqrt{3}%
\pi f_{\pi }e_{2}\hat{u}}{\beta _{0}}\int_{0}^{\infty
}due^{-t_{2}u}B[u,\alpha _{s}(-\hat{t}/2)].  \label{eq:16}
\end{equation}%
Here%
\begin{equation*}
t_{1}=\frac{4\pi }{\beta _{0}\alpha _{s}(-\hat{u})},\;t_{2}=\frac{4\pi }{%
\beta _{0}\alpha _{s}(-\hat{t})},
\end{equation*}%
and functions $A[u,\alpha _{s}(-\hat{u}/2)]$ , $B[u,\alpha _{s}(-\hat{t}/2)]$
contain IR renormalon poles, number and locations of which depend on the
meson (pion) distribution amplitude. These functions depend also on the
factorization scales through $\alpha _{s}(-\hat{u}/2)$ and $\alpha _{s}(-%
\hat{t}/2)$, respectively. In other words correct treatment leads to
expressions with two scales and two integrals, but not to formulas like
(3.10)-(3.13) in Ref.\cite{Az2}%
\begin{equation*}
D(Q^{2})=\left( \frac{4\sqrt{3}\pi f_{\pi }e_{1}\hat{t}}{\beta _{0}}+\frac{4%
\sqrt{3}\pi f_{\pi }e_{2}\hat{u}}{\beta _{0}}\right)
\int_{0}^{\infty }due^{-tu}A[u,\alpha _{s}(p_{T}^{2})],
\end{equation*}%
with%
\begin{equation*}
t=\frac{4\pi }{\beta _{0}\alpha _{s}(p_{T}^{2})}.
\end{equation*}%
Therefore Eqs.(3.10)-(3.13) in Ref. \cite{Az2} are wrong. Authors state that
they found the relevant Borel transforms (3.10)-(3.13) of the resummed
expressions . In the case of correct choice for the scales Eqs.(3.14)-(3.17)
are wrong, because as it follows from our analysis and from Eq.(\ref{eq:16})
the corresponding Borel transforms have two components:%
\begin{equation}
B_{1}[D(\hat{t},\hat{u})](u)=A[u,\alpha _{s}(-\hat{u}/2)],\;\;\;B_{2}[D(\hat{%
t},\hat{u})](u)=B[u,\alpha _{s}(-\hat{t}/2)].  \label{eq:17}
\end{equation}%
As a result, the resummed expressions should have also two pieces. Then, for
example, for Eq.(\ref{eq:16}) we get
\begin{equation}
\label{eq:18}
\lbrack D(\hat{t},\hat{u})]^{res}=\frac{4\sqrt{3}\pi f_{\pi }e_{1}\hat{t}}{%
\beta _{0}}\tilde{A}[Li(\lambda _{1}^{n})/\lambda _{1}^{n},\alpha _{s}(-%
\hat{u}/2)]+\frac{4\sqrt{3}\pi f_{\pi }e_{2}\hat{u}}{\beta _{0}}\tilde{B}%
[Li(\lambda _{2}^{n})/\lambda _{2}^{n},\alpha _{s}(-\hat{t}/2)],
\end{equation}%
where%
\begin{equation*}
\lambda _{1}=\frac{-\hat{u}}{\Lambda ^{2}},\;\;\lambda _{2}=\frac{-\hat{t}}{%
\Lambda ^{2}},
\end{equation*}%
and the logarithmic integral $Li(\lambda )$ for $\lambda >1$ is defined in
its principal value%
\begin{equation*}
Li(\lambda )=P.V.\int_{0}^{\lambda }\frac{dx}{\ln x}.
\end{equation*}%
(in Eq.(3.22) of Ref.\cite{Az2} there is evident error/misprint).
Consequently, Eqs.(3.18)-(3.21) presented in Ref.\cite{Az2} are
wrong as well. Our analysis of the Borel transforms and the
resummed expressions presented in Eqs. (\ref{eq:15})-(\ref{eq:18})
are valid also for the first process $\gamma+\gamma \rightarrow
M+X$ \cite{Az1}.

We think that after this analysis the remaining part of the works \cite{Az1}
and \cite{Az2}, namely the leading twist contribution to the process and
numerical calculations do not deserve further discussions. Unfortunately,
all these fatal errors, wrong statements, numerous misprints were overlooked
by referees of such prestigious journal like Phys. Rev. D. Therefore, we
decided to inform the HEP community on ''methods'' used by these authors in
their ''investigations''.

\end{document}